\documentclass[mathleft
]{an}
\usepackage{graphicx}
\usepackage{times}
\usepackage{amsmath,amssymb}
\begin{document}

% The following seven commands are intended for editorial usage and should be ignored by
% the author(s).
\Pagespan{1}{}% Document's page range. 
% If second parameter is left empty, the last page is computed automatically.
\Yearpublication{2010}%
\Yearsubmission{2010}%
\Month{04}%   
\Volume{0000}%  
\Issue{00}% 
% \DOI{This.is/not.aDOI}% 

\def\note #1]{{\bf #1]}}

\title{Convection and oscillations}

%\author{G. Houdek\inst{1}\fnmsep\thanks{Corresponding author:
\author{G. Houdek\thanks{%Corresponding author:
  \email{guenter.houdek@univie.ac.at}\newline}
%Example 
%for footnote, note the usage of the \texttt{fnmsep}
%command as separator between institute number and footnote mark} 
%\and  G.H. Ostwriter\inst{2,3}
}
\titlerunning{Convection and oscillations}
\authorrunning{G. Houdek}
\institute{
Institute of Astronomy, University of Vienna, 
A-1180 Vienna, Austria}

\received{??}
\accepted{??}
\publonline{later}


\abstract{%
  In this short review on stellar convection dynamics I address the following,
  currently very topical, issues:
  (1) the surface effects of the Reynolds stresses and nonadiabaticity on solar-like
  pulsation frequencies, and (2) oscillation mode lifetimes of stochastically
  excited oscillations in red giants computed with different time-dependent 
  convection formulations.
  }

\maketitle

\section{Introduction}
Seismic data of solar-type oscillations in distant stars are being provided 
with high precision by the space missions CoRoT (Baglin et al. 2009) and Kepler 
(Borucki et al. 2009; Christensen-Dalsgaard 2007), and hopefully also soon
from the ground-based Danish SONG project (Grundahl et al. 2010). 
In contrast to solar observations the observed oscillation modes in distant 
stars are of only low spherical degree $l$. However, 
these are the modes which provide information from the deepest layers of the star.
Some theoretically important properties, such as the gross structure of the 
energy-generating core and the extent to which it is convective, and the 
large-scale variation of the angular velocity, will become available. 
Such information will be of crucial importance for checking, and then 
calibrating, the theory of the structure and evolution of stars.

Solar-type oscillations are characterized by being intrinsically damped and
stochastically driven by the near-surface turbulent velocity field 
(e.g., Goldreich \& Keeley 1977, Balmforth 1992b, Houdek et al. 1999, Samadi et al. 2008). 
The turbulent convection produces acoustic noise in a broad frequency range, thereby
exciting many of the global solar-type oscillations to observable amplitudes. 
The resulting rich acoustic
power spectrum makes these modes particularly interesting for studying fluid-dynamical
aspects of the stellar interior, and it is our hope that their mode properties can be
used to probe the still ill-understood near-surface regions.
The physics of these near-surface regions is, besides many other effects, further 
complicated by the dynamics of the transport of the turbulent fluxes of heat and momentum
and their coupling to the pulsations. Several attempts have been made in the past to
model the effects of convection dynamics on the oscillation frequencies, for which
a time-dependent convection formalism is required.
The influence of the momentum flux and its pulsationally 
induced perturbation upon adiabatic and nonadiabatic pulsation 
frequencies were investigated first by Gough (1984), using a local
time-dependent mixing-length formulation for convection (Gough 1977a), but
a simplified analytical approximation to the eigenfunctions 
in the atmosphere. 
Gough concluded that the inclusion of the momentum flux in the mean model 
reduces the frequency residuals between observations and model computations, 
whereas the additional dynamical effects through their pulsationally induced 
perturbations actually aggravates this frequency discrepancies. Balmforth (1992a) 
studied these effects in a more consistent way and by means of the more sophisticated 
nonlocal time-dependent mixing-length formulation by Gough (1977b).
Balmforth also concluded that the modification of 
the mean stratification by the momentum flux 
substantially decreases the adiabatic frequency residuals, whereas the effects of 
the momentum flux perturbation and nonadiabaticity lead to an increase in the 
modelled oscillation frequencies. 

In Section 2 we discuss the results of more recent attempts to model the effects of
the near-surface layers on the oscillations frequencies in the Sun and other stars. 
The other topical issue on the comparison between observed p-mode lifetimes in red giants and
theoretical predictions, obtained with different time-dependent convection models, is
addressed in Section 3.

\begin{figure}[t]
\centering
\includegraphics[width=0.38\textwidth]{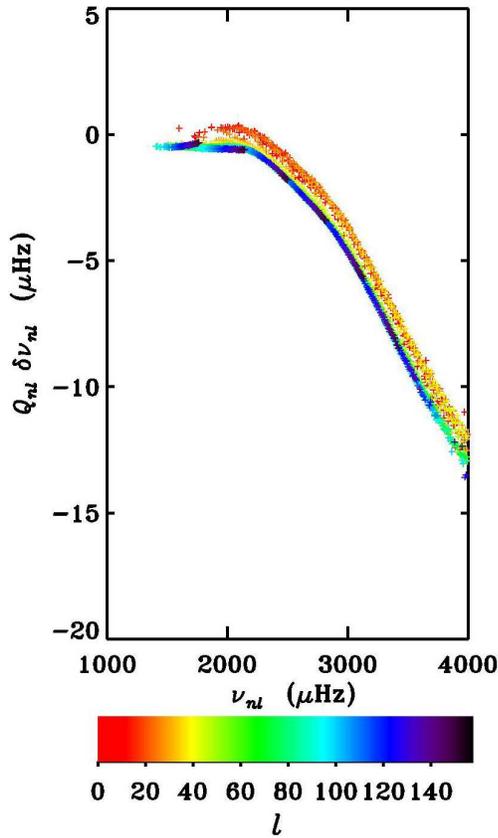}
\caption{Scaled differences between observed GONG frequencies and
         adiabatically computed frequencies of the `standard' solar Model S.
         The degree $l$ of the oscillation modes is indicated by the colour bar
         (from Christensen-Dalsgaard et al. 1996).}
\label{fig:gong-ModelS}
\end{figure}

\begin{figure}[t]
\centering
\includegraphics[width=0.37\textwidth]{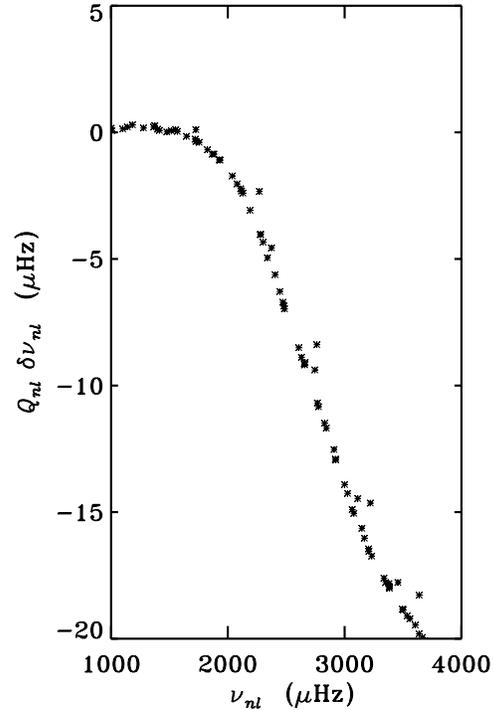}
\caption{Scaled adiabatic frequency differences between a model
         for which the near-surface layers were represented
         by a hydrodynamical simulation, and the `standard' 
         solar Model S (from Rosenthal et al. 1995).}
\label{fig:rosenthal}
\end{figure}

\section{The effect of convection dynamics on the oscillation frequencies}
The near-surface layers in the Sun and other stars are still poorly understood,
mainly because of the rather difficult to model complex physics of the interaction
of convection, radiation, magnetism and rotation. Typical stellar structure 
calculations treat these superficial layers with simplified atmosphere models 
and the stratification of the superadiabatic region by means of a mixing-length 
approach (e.g. B\"ohm-Vitense 1958). Furthermore, the effects of the mean Reynolds
stresses on the hydrostatic equilibrium are essentially always ignored.
Linear pulsation calculations typically adopt the adiabatic approximation and
ignore the momentum flux perturbations (turbulent pressure fluctuations).
However, nonadiabatic effects and the fluctuations of the turbulent fluxes
(heat and momentum) do modify the modelled pulsation eigenfunctions and
consequently also the oscillation frequencies (e.g., Gough 1980, 
Balmforth 1992b, Houdek 1996).\\
These effects are predominantly confined to the upper-most stellar layers,
where the vertical scale of low- or intermediate degree modes is much less 
than the horizontal scale and when $l$ is low the mode inertia is quite
insensitive to degree $l$. Consequently the observed oscillation properties 
of low-degree modes depend little on the value of $l$ (e.g., Libbrecht 1986).
Moreover, the influence of the near-surface effects on the oscillation
frequencies of low-degree modes is predominantly a function of frequency alone
(e.g., Christensen-Dalsgaard \& Thompson 1997), whereas for modes with
intermediate degree a weak degree dependence is observed. This degree dependence,
however, can be described by a simple scaling $Q_{nl}=I_{nl}/\overline I_0$ with mode 
inertia $I_{nl}$ (e.g., Aerts, Christensen-Dalsgaard \& Kurtz 2010, $\S$7.1.4), where 
$n$ is the radial mode order, and $\overline I_0(\nu)$ is obtained from interpolating the 
radial mode inertia $I_{n0}$ to the nonradial oscillation frequency $\nu_{nl}$.
For a `standard' solar model, such as Model S (Christensen-Dalsgaard et al. 1996), 
the (scaled) differences between the solar and model frequencies are indeed 
dominated by the near-surface effects and are predominantly a function of
frequency alone. Differences between frequencies observed
with the GONG instruments and adiabatically computed Model S frequencies are 
illustrated in Fig.\,\ref{fig:gong-ModelS}.
The increase of the frequency residuals with oscillation frequency depends on 
the modelling details of the functional form of the acoustic cutoff frequency 
$\nu_{\rm ac}$ (as it affects the acoustic potential) with radius in the near-surface 
layers. For an isothermal atmosphere $\nu_{\rm ac}=c/4\pi H$, where $c$ is the adiabatic 
sound speed and $H$ the pressure scale height.
In the Sun $\nu_{\rm ac}\simeq5.5\,$mHz.
Thus, we see from Fig.\,\ref{fig:gong-ModelS} that the influence of the near-surface 
layers is most important, when the oscillation frequency $\nu$ is comparable to 
the acoustic cutoff frequency $\nu_{\rm ac}$. The cutoff frequency determines the
location at which an incident acoustic wave is reflected back into the stellar interior, 
and the lower the frequency $\nu$, the deeper the location at which this reflection takes
place. For modes with frequencies $\nu$ much less than $\nu_{\rm ac}$ reflection takes 
place so deep
in the star that the modes are essentially unaffected by the near-surface structure.
When $\nu$ is comparable with $\nu_{\rm ac}$, however, the inertia of the near-surface
layers is a considerable fraction of the total mass above the reflecting layer, leading
to a greater modification to the phase shift in the spatial oscillation eigenfunctions,
and, through the dispersion relation, also to a change in frequency. The inertia of the 
essentially hydrostatically moving near-surface layers depends on mass and consequently 
on the equilibrium pressure near the photosphere. 

\subsection{The effect of the Reynolds stresses in the mean structure}
\label{sec:pt-in-mean_model}
\begin{figure}[t]
\centering
\includegraphics[width=0.48\textwidth]{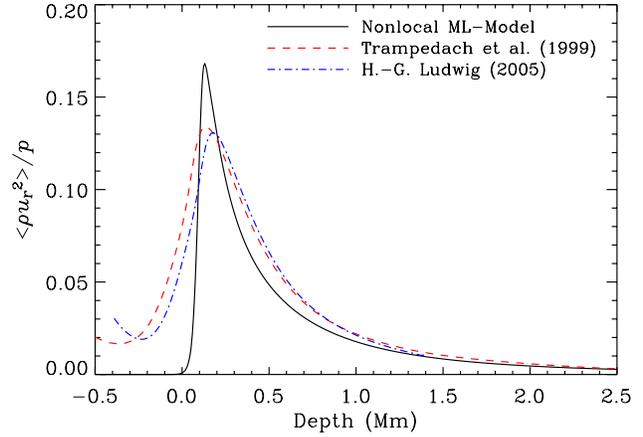}
\caption{Turbulent pressure $p_{\rm t}=\langle\rho u^2_r\rangle$ over total pressure 
$p=p_{\rm g}+p_{\rm t}$ as a function of the depth variable R$_\odot-r$ for 
various solar simulations and models (R$_\odot$ is the photospheric solar radius). 
Results are shown for the nonlocal 
mixing-length model by Gough (1977ab, solid curve) and from hydrodynamical 
simulations by Trampedach (1999, personal communication, dashed curve) and 
Ludwig (2005, personal communication, dot-dashed curve).}
\label{fig:pt_sun}
\end{figure}

From the discussion before we conclude that the details of modelling the 
hydrostatic equilibrium structure in the near-surface layers play an important 
role in describing the residuals between observed and modelled oscillation
frequencies, particularly for modes with $\nu$ close to $\nu_{\rm ac}$. 
Almost all stellar model calculations consider only the gradient of the
gas pressure $p_{\rm g}$ in the equation of hydrostatic support. In the
convectively unstable surface layers, however, the turbulent velocity 
field $\boldsymbol{u}$ contributes to the hydrostatic support
via the Reynolds stresses $\langle\rho\boldsymbol{uu}\rangle$ (angular brackets 
denote an ensemble average), the $(r,r)$ component
$p_{\rm t}:=\langle\rho{u_ru_r}\rangle$ of which, acts as a pressure
term additionally to the gas pressure $p_{\rm g}$ (e.g., Gough 1977ab).
Solar models indicate that the turbulent pressure $p_{\rm t}$ can be as
large as 15\% of the total pressure $p=p_{\rm g}+p_{\rm t}$, as 
illustrated in Fig.\,\ref{fig:pt_sun} for three different 
simulations and models of the Sun.

Hydrodynamical simulations of stellar convection have enabled us to estimate the
turbulent pressure (see Fig.\,\ref{fig:pt_sun}). Rosenthal et al. (1995) investigated
the effect on adiabatic eigenfrequencies of the contribution that the turbulent 
pressure makes to the mean hydrostatic stratification: he examined a hydrodynamical 
simulation by Stein \& Nordlund (1991) of the outer $2\%$ by radius of the Sun, matched 
continuously in sound speed to a model envelope calculated, as in a `standard' solar model, 
with a local mixing-length formulation. 
The resulting frequency shifts of adiabatic oscillations between the simulations and the
`standard' solar reference model, Model S, are illustrated in Fig.\,\ref{fig:rosenthal}. 
The frequency residuals behave similarly to the data in Fig.\,\ref{fig:gong-ModelS}, though
with larger shifts at higher oscillation frequencies.

If turbulent pressure is considered in the mean structure, 
additional assumption have to be made about the turbulent pressure
perturbation in the adiabatic pulsation equations. The first adiabatic exponent 
$\gamma_1=(\partial\ln p_{\rm g}/\partial\ln\rho)_s$ ($s$ being the specific entropy), 
is a purely thermodynamic quantity
%, and which is for an ideal gas proportional
%to the ratio of the two specific heats at constant (gas) pressure $c_{p_{\rm g}}$ 
%and volume $c_V$, 
and is expressed by means of the gas pressure $p_{\rm g}$. 
Consequently, in the presence of turbulent pressure 
$p_{\rm t}$, such that the total pressure $p=p_{\rm g}+p_{\rm t}$ 
satisfies the equation of hydrostatic equilibrium, % [equation (\ref{eq:hydrostat})], 
$\gamma_1$ experiences a modification of the form
\begin{equation}
\tilde{\gamma}_1=\left(\frac{\partial\ln p}{\partial\ln\rho}\right)_{\!\!s}=
\frac{1}{p}\left[\left(\frac{\partial p_{\rm g}}{\partial\ln\rho}\right)_{\!\!s}+
           \left(\frac{\partial p_{\rm t}}{\partial\ln\rho}\right)_{\!\!s}
          \right]
\label{eq:gamma_tilde}
\end{equation}
%in the linearized adiabatic continuity equation (Rosenthal et al. 1995).
and the Lagrangian perturbation in the total pressure is
\begin{equation}
\frac{\delta p}{p}=\frac{\delta p_{\rm g}}{p}+
\frac{\delta p_{\rm t}}{p}=
\tilde{\gamma}_1\,\frac{\delta\rho}{\rho}
\,.
\label{eq:adiabat-press-fluc}
\end{equation}
Nonadiabatic pulsation calculations with
the inclusion of the turbulent pressure fluctuation $\delta p_{\rm t}$ (Houdek 1996), 
and hydrodynamical simulation results (Rosenthal et al. 1995) indicate
that $\delta p_{\rm t}$ varies approximately in quadrature 
with the other terms in the linearized momentum equation, and hence contributes
predominantly to the imaginary part of the frequency shift, i.e. to the 
linear damping rate. The Lagrangian perturbation $\delta p_{\rm g}$, however, 
responds adiabatically. Therefore $\delta p_{\rm t}$ can be neglected
in equation (\ref{eq:adiabat-press-fluc}), i.e in the calculation of the 
real adiabatic eigenfrequencies it is assumed that 
$\delta p/p\simeq\delta p_{\rm g}/p\simeq\tilde\gamma_1\delta\rho/\rho$. With this assumption
$\tilde{\gamma}_1\simeq(p_{\rm g}/p)\gamma_1$, and the only modification to 
the adiabatic oscillation equations is the replacement of $\gamma_1$ by $\tilde\gamma_1$
(Rosenthal et al. 1995).

\vspace{-5pt}
\subsection{The effects of nonadiabaticity and momentum flux perturbation}
The effects of nonadiabaticity and convection dynamics on the pulsation 
frequencies were, for example, studied by Balmforth (1992b), Rosenthal et al. (1995) 
and Houdek (1996). In these studies the nonlocal, time-dependent generalization 
of the mixing-length formulation by Gough (1977ab) was adopted to model the heat 
and momentum flux consistently in both the equilibrium envelope model and in the 
nonadiabatic stability analysis. Houdek (1996) considered the following models:
\begin{quote}
\begin{itemize}
\item[\hfill L.a] A local mixing-length formulation without turbulent pressure $p_{\rm t}$
           was used to construct the mean envelope model. Frequencies were
           computed in the adiabatic approximation assuming $\delta p_{\rm t}=0$
           (see Section\,\ref{sec:pt-in-mean_model}).
\item[\hfill NL.a] Gough's (1977ab) nonlocal, mixing-length model, including turbulent
            pressure, was used to construct the mean envelope model. 
            Frequencies were computed in the adiabatic approximation assuming 
            $\delta p_{\rm t}=0$.
\item[\hfill NL.na] The mean envelope model was constructed as in NL.a. 
             Nonadiabatic frequencies were computed including consistently the 
             Lagrangian perturbations of the convective heat flux $\delta F_{\rm c}$
             and turbulent momentum flux $\delta p_{\rm t}$.
\end{itemize}
\end{quote}
Additional care was necessary when frequencies between
models with different convection treatments were compared, such as in the models L.A and NL.a.
In order to isolate the effect of the near-surface structures on the oscillation frequencies
the models had to posses the same stratification in their deep interiors. This was 
obtained by requiring the models to lie on the same adiabat near the base of the
(surface) convection zone and to have the same convection-zone depth.
Varying the mixing-length parameter $\alpha=H/\ell$ ($\ell$ is the mixing length)
and hydrogen abundance by iteration in model L.a, the same values for 
temperature and pressure were found at the base of the convection zone than those in
models NL.a and NL.na. The radiative interior of the nonlocal models NL.a and NL.na were 
then replaced by the solution of the local model L.a, and the convection-zone depth was
calibrated to $0.287$~R$_\odot$ (Christensen-Dalsgaard, Gough \& Thompson 1991).
Further details of the adopted physics 
in the model calculations can be found in Houdek et al. (1999).

\begin{figure}[t]
\centering
\includegraphics[width=0.45\textwidth]{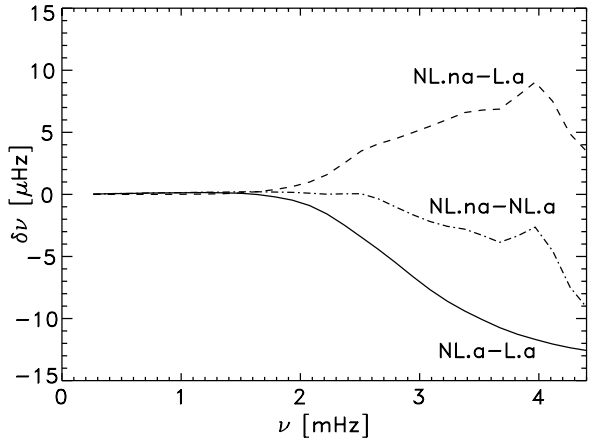}
\caption{Frequency residuals between solar models calculated with a nonlocal
generalization and the 'standard' local mixing-length formulation.
The solid curve (NL.a-L.a) shows the adiabatic radial oscillation frequency shifts, under
the assumption $\delta p_{\rm t}=0$, caused by the turbulent pressure $p_{\rm t}$ in the mean
envelope. The dashed curve (NL.na-NL.a) is the frequency shift caused by nonadiabaticity 
and effects of including consistently $\delta p_{\rm t}$.
The overall frequency shift (NL.na-L.a) is plotted by the dot-dashed curve
(from Houdek 1996).}
\label{fig:NL_frequ-diff_sun}
\end{figure}

The outcome of these calculations is shown in Fig.\,\ref{fig:NL_frequ-diff_sun}.
As for the hydrodynamical simulations (Fig.\,\ref{fig:rosenthal}) the effect of the
Reynolds stresses in the mean structure decreases the adiabatic frequencies 
(NL.a-L.a, solid curve) for frequencies larger than about 2\,mHz, though the maximum deficit
of about 12$\,\mu$Hz is smaller than in the hydrodynamical simulations. The effects
of nonadiabaticity and $\delta p_{\rm t}$ (NL.na-L.a, dashed curve), however, lead to an 
increase of the mode frequencies by as much as $\sim 9\,\mu$Hz, nearly cancelling 
the downshifts from the
effect of $p_{\rm t}$ in the mean structure, as illustrated by the dot-dashed 
curve (NL.na-NL.a).
If the positive frequency shifts between models NL.na and L.a (dotted curve)
are interpreted as the nonadiabatic and momentum flux corrections to the oscillation
frequencies then their effects are to bring the frequency residuals of the hydrodynamical 
simulations (Fig.\,\ref{fig:rosenthal}) in better agreement with the data plotted in
Fig.\,\ref{fig:gong-ModelS}. The effects of the near-surface regions in the Sun were
also considered by Rosenthal et al. (1999) and Li et al. (2002) based on hydrodynamical
simulations. 

\begin{figure}[t]
\centering
\includegraphics[width=0.41\textwidth]{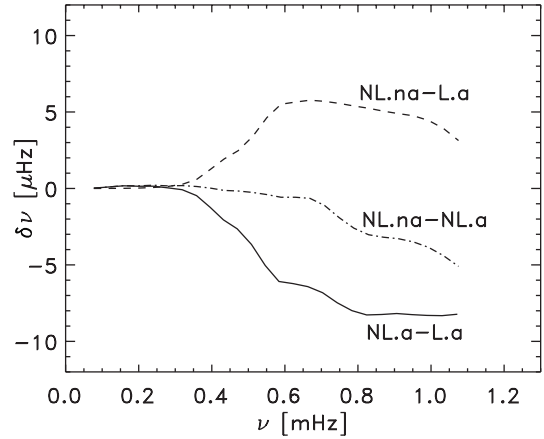}
\caption{Frequency residuals between models for the solar-like star $\eta\,$Boo.
The model calculations and line styles are as in Fig.\,\ref{fig:NL_frequ-diff_sun}
(from Houdek 1996).}
\label{fig:frequ-diff_etaBoo}
\end{figure}

A similar conclusion as in the solar case was also found for the solar-like star $\eta\,$Boo 
by Christensen-Dalsgaard et al. (1995), Houdek (1996), demonstrated in 
Fig.\,\ref{fig:frequ-diff_etaBoo}, and more recently by Straka et al. (2006).

The near-surface frequency corrections also affect the determination of the modelled 
mean large frequency separation $\Delta\nu:=\langle\nu_{n+1l}-\nu_{nl}\rangle$ (angular 
brackets indicate an average over $n$ and $l$). In both models for the Sun and for 
$\eta\,$Boo the resulting
corrections to $\Delta\nu$ are about -1$\,\mu$Hz. Although this correction is less than 1\% it
does affect the determination of the stellar radii and ages from the observed values of
$\Delta\nu$ and small frequency separation $\delta\nu_{02}$ in distant stars. 
A simple procedure for estimating the near-surface frequency corrections was suggested recently
by Kjeldsen et al. (2008), based on the {\it ansatz} that the frequency shifts can be scaled
as $a(\nu/\nu_0)^b$ (Christensen-Dalsgaard \& Gough 1980), where $\nu_0$ is a suitable reference 
frequency, $a$ is determined from fitting this expression to the observed frequencies, 
and $b$ is obtained from the solar data (see also Christensen-Dalsgaard, these proceedings).  
It is to be hoped that the analyses of high-quality data for a broad range of stars,
observed from the space mission Kepler, will contribute to a better understanding of
the near-surface effects.

\begin{figure}[t]
\centering
\includegraphics[width=0.45\textwidth]{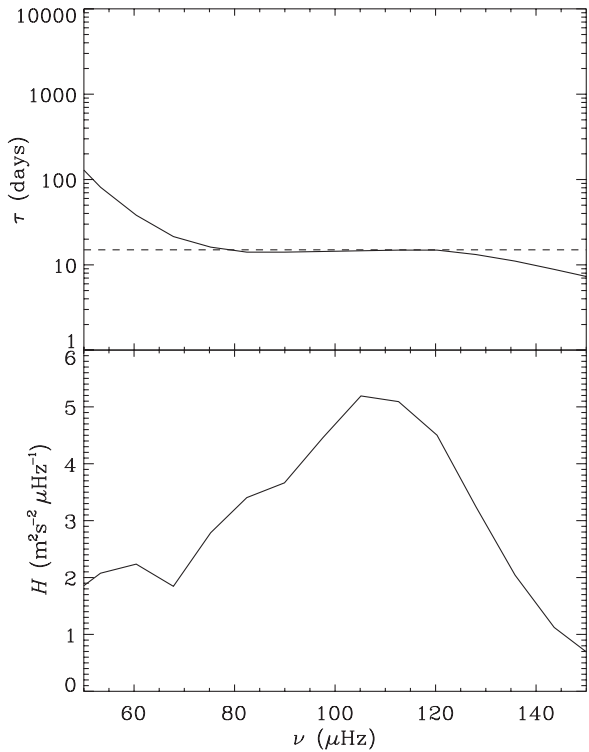}
\caption{Radial pulsation mode properties of solar-type oscillations for a model of 
         the red-giant star $\xi\,$Hydrae. The computations adopted Gough's (1977)
         time-dependent convection model.
         Top: theoretical mode lifetimes as a function of pulsation frequency. The 
         horizontally dashed line indicates a mode lifetime of 15 days.
         Bottom: modelled oscillation mode heights as a function of pulsation frequency
         (from Houdek \& Gough 2002).
        }
\label{fig:xiHya}
\end{figure}

\begin{figure}[t]
\centering
\includegraphics[width=0.465\textwidth]{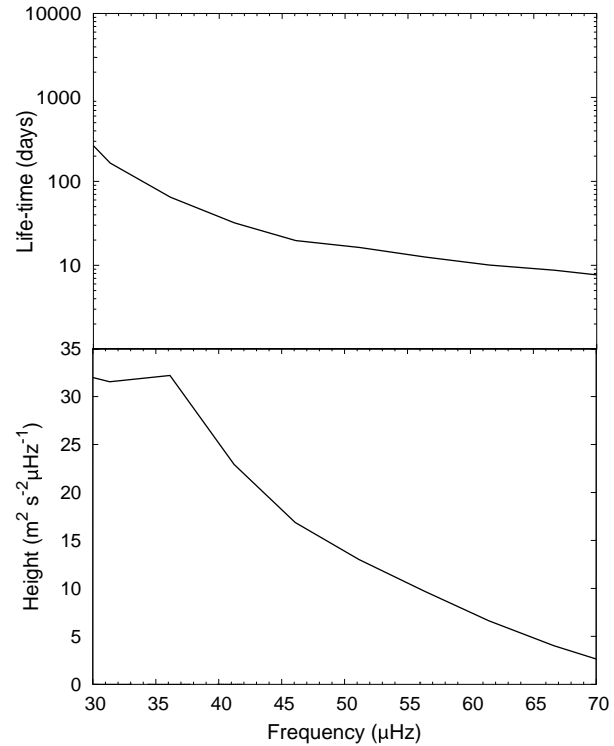}
\caption{Radial pulsation mode properties of solar-type oscillations for
         a red-giant model in the core helium-burning phase. The computations
         were carried out by Dupret et al. (2009) using the convection model
         by Grigahc\`ene et al. (2004).
         Top: theoretical mode lifetimes as a function of pulsation frequency. The 
         radial mode lifetimes decrease monotonically with frequency.
         Bottom: modelled oscillation mode heights as a function of pulsation frequency
         (adopted from Dupret et al. 2009).
        }
\label{fig:Dupret_ModelB}
\end{figure}

\section{Mode lifetimes in red giants}
The first convincing detection of solar-type oscillations in a red-giant star
was announced by an international team of astronomers (Frandsen et al. 2002) for
the star $\xi\,$Hydrae (G7III) with velocity oscillation amplitudes of about 2\,ms$^{-1}$. 
%In some respects its structure differs more from that of the Sun than does that of any of the
%other stars in which such oscillations have been seen, and one might therefore expect it to
%provide a more stringent test of oscillation properties than have been estimated by theoretical
%scaling from solar conditions. 
Houdek \& Gough (2002) calculated mode properties 
for the red-giant star $\xi\,$Hydrae and reported velocity amplitudes that were in good 
agreement with the observations. Moreover, using Gough's (1977ab) time-dependent convection 
model, these authors estimated theoretical mode lifetimes and reported for the most prominent
p modes a lifetime $\tau$ of about 15--17 days (see Fig.\,\ref{fig:xiHya}). 
This prediction was, however, later challenged by Stello et al. (2004, 2006), who 
developed a new method for measuring mode lifetimes from various properties of the 
observed oscillation
power spectrum and reported a measured mode lifetime of only 2-3 days for the 
star $\xi\,$Hydrae. 
This is in stark contrast to the predicted values by Houdek \& Gough (2002), a discrepancy that 
needs to be understood. In the new method by Stello et al. it is
assumed that the observed oscillation power spectrum is dominated by
radial modes, because nonradial modes have larger inertiae $I$ and consequently lower 
amplitudes $V$ (e.g. Dziembowski et al., 2001, Christensen-Dalsgaard 2004).
The maximum peaks in the observed power spectrum, however, which are actually mode heights $H$
and related to the velocity amplitudes $V$ via the relation $V^2=\eta H/2$ (e.g.,
Chaplin et al. 2005), in which $\eta=\tau^{-1}$ is the damping rate in units of 
angular frequency, are independent of the pulsation mode inertiae. 
Consequently the pulsation heights $H$ of nonradial modes could have similar values
to those of radial modes of similar frequencies, provided the modes are 
resolved (e.g. Dupret et al. 2009).
A first hint, shedding some light on the pulsation mode lifetimes in red giants, has been 
provided by oscillation data obtained by CoRoT (e.g., De Ridder et al. 2009) from several 
red-clump stars, and by Kepler (Bedding et al. 2010). These data support the 
possibility of the presence of nonradial pulsation modes (see also Hekker, these proceedings), 
bringing observed mode lifetimes and theoretical predictions in better agreement (see also
Carrier et al. 2010).

\noindent Fig.\,\ref{fig:xiHya} illustrates the predictions of radial mode lifetimes 
and oscillation
heights for a model of the red giant star $\xi\,$Hydrae. As in the Sun the predicted mode 
lifetimes in $\xi\,$Hydrae show a pronounced plateau about the maximum pulsation height
near 105$\,\mu$Hz, a frequency value that is in good agreement with the observations
by Frandsen et al. (2002).\\
Red-giant stars may display different 
frequency patterns in the oscillation power spectrum, from regularly to
very complicated patterns, depending on the density contrast between the core
and the envelope (e.g., Dupret et al. 2009).
Evolutionary calculations suggest that $\xi\,$Hydrae is most likely in the 
core helium-burning phase (Teixeira et al. 2003; Christensen-Dalsgaard 2004). 
Dupret et al. (2009) computed mode lifetimes and pulsation heights for
several red-giant models in different evolutionary phases. 
Mode properties of radial pulsations for a red-giant model in the core 
helium-burning phase, calculated by Dupret et al., are displayed in 
Fig.\,\ref{fig:Dupret_ModelB}.
Although the results in Figs\,\ref{fig:xiHya} and \ref{fig:Dupret_ModelB} are for
different models, though both models are in the core helium-burning phase, it is 
here assumed that their pulsation properties can be compared qualitatively. The 
most interesting
qualitative difference between Figs\,\ref{fig:xiHya} and \ref{fig:Dupret_ModelB}
is the frequency dependence of the radial mode lifetime 
(top panels) showing a monotonic decrease with frequency 
in the results by Dupret et al.  This could be a result of having adopted 
different convection formulations in the model computations. 
Also, the frequency dependence of the pulsation mode heights (bottom panels) are different.
In the stability computations by Dupret et al. (Fig.\,\ref{fig:Dupret_ModelB}) the 
turbulent fluxes were modelled according to the time-dependent 
convection formulation by Grigahc\`ene et al. (2004), which is based on Unno's (1967) 
formulation. Houdek \& Gough (2002) adopted Gough's (1977ab) time-dependent nonlocal 
convection formulation in the model computations for $\xi\,$Hydrae (Fig.\,\ref{fig:xiHya}).
Stability computations of red-giant oscillations were also recently addressed by 
Xiong \& Deng (2007) using Xiong's (1989) time-dependent convection formulation.
Some of the model differences between Unno's (1967) and Gough's (1977a)
convection formulation were discussed by Houdek (2008).
It is to be hoped that the high-quality data from, for example Kepler, will not 
only help modellers to calibrate but also to improve convection models for pulsating stars.
 
\acknowledgements
I thank the referee for his helpful comments.
Support by the Austrian Science Fund (FWF project P221205) is 
thankfully acknowledged.    

%\newpage%%%%%%%%%%%%%%%%%%%%%%%%%%%%%%%%%%%%%%%%%%%%%%%%%%%%%%

\end{document}